\documentclass[
 twocolumn,
  prl,
  aps,
  superscriptaddress,
  showpacs,preprintnumbers,
  amsmath,amssymb,
floatfix,
A4,
]{revtex4-1}

\usepackage{amsmath}
\usepackage{amssymb}
\usepackage{graphicx,color}
\usepackage{multirow}

\begin{document}

% Title of the article
\title{Conductance fluctuations in graphene devices with super\-conducting contacts in different charge density regimes}

% Authors
\author{Frank Freitag}
\author{Jelena Trbovic}
\author{Christian Sch\"onenberger}
 \email{Christian.Schoenenberger@unibas.ch}
\affiliation{Departement Physik, Universit\"at Basel, Klingelbergstr. 82, 4056 Basel, Switzerland}

\begin{abstract}
    Conductions fluctuations (CF) are studied in single layer graphene devices
    with super\-conducting source and drain contacts made from aluminium. The CF are found to
    be enhanced by super\-conductivity by a factor of 1.4 to 2. This (near) doubling of the CF indicates
    that the phase coherence length is $l_{\phi} \gtrsim L/2$.
    As compared to previous work, we find a relatively weak dependence of the CF on the gate
    voltage, and hence on the carrier density. We also demonstrate that whether the CF are larger
    or smaller at the charge neutrality point can be strongly dependent
    on the series resistance $R_{C}$, which needs to be subtracted.
\end{abstract}

\maketitle  

\section{Introduction}

Graphene has proven to be interesting for the study of conductance
fluctuations (CF) in disordered systems
\cite{Morozov2006,Heersche2007,Ojeda-Aristizabal2009,Skakalova2009,Trbovic2010}.
Some of its remarkable properties are the linear low energy band
structure \cite{Geim2007} and the K-K' degeneracy, which leads to
unusual scattering \cite{McCann2006}. In common supported graphene
devices disorder is inevitably introduced by the substrate and CF
have been shown to be major corrections to the transport at low
temperature.

In numerical calculations \cite{Rycerz2007,Borunda2011} stronger CF were found away from the charge neutrality point (CNP). Additionally, the CF were reported to be non-ergodic and they were suspected to result from variations in trajectories, rather than in phase shifts \cite{Rycerz2007}. The results and interpretation of \cite{Rycerz2007} were debated in \cite{Kharitonov2008}, where CF were studied analytically and a similar magnitude of the CF was found but with fulfilled ergodicity.

Superconducting electrodes can provide additional information on the
transport  mechanisms in disordered systems \cite{Beenakker1995}.
Phase sensitive Andreev reflection can occur at the interface
between the super\-conductor (S) and the normal conducting graphene
(G). At energies below the super\-conducting gap $\Delta$ electrons
from G can only enter S in the form of a Cooper pair. This charge
transfer needs time reversal symmetry, which can be broken by the
application of an external magnetic field.

In the following, we look at CF in a two-terminal graphene device
with  super\-conducting contacts. We compare the CF in the
super\-conducting and normal state of the electrodes. The latter is
achieved by either increasing the temperature or by applying a
magnetic field. The CF near and far away from the CNP are presented.

%-------------------------------------

\section{Sample preparation and experimental set-up}
Mechanical Microcleavage \cite{Novoselov2005} was used to deposit
the  graphene flakes onto a highly p-doped silicon wafer, covered by
300~nm silicon dioxide. The silicon wafer serves as a back-gate
where a voltage $V_{gate}$ is applied.

The graphene flakes on the substrate were located with optical
microscopy and  categorised by optical contrast. To ensure that the
selected flakes are indeed single layer graphene, Raman spectroscopy
\cite{Ferrari2006} with 532~nm laser light was performed.

Subsequently, a PMMA mask was used to pattern the electrical lines
by  electron beam lithography. A thin metal film (5~nm Ti / 50~nm
Al) was evaporated on the sample in a UHV e-gun evaporation system.
After lifting the PMMA mask off, the sample was cleaned in chemical
solvents (acetone, ethanol and 2-propanol) and annealed in vacuum at
200~$^\circ$C for several hours.

The inset of figure \ref{BG and SEM} shows a micrograph of the
device  investigated. The single layer graphene flake is
approximately of length $L=$~1.3~$\mu$m and width $W=$~3.8~$\mu$m.

Electrical measurements were performed in an Oxford Instruments
$^3$He  cryostat at temperatures in the range of 230~mK - 1.5~K. A
Stanford Research Systems SR830 lock-in amplifier was used to apply
a 20~$\mu$V AC voltage which was superimposed onto a DC voltage
$V_{sd}$. The current through the sample was detected by an IV
converter. No series resistors (like the electrical lines of the
cryostat or the impedance of the IV converter) were subtracted from
the data, unless mentioned.

%-------------------------------------

\section{Results and discussion}

\begin{figure}[h!]%
\center
\includegraphics{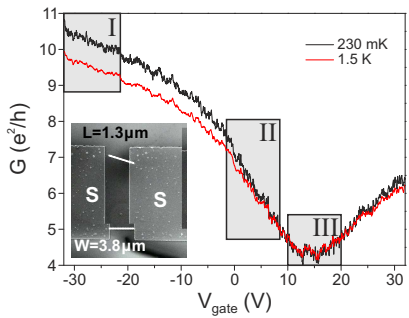}
\caption{(online colour at: www.pss-b.com) The response of the
conductance $G$ through the single  layer graphene device on the
applied back-gate voltage $V_{gate}$ at 230~mK and 1.5~K (lower
curve). The charge neutrality region lies at $V_{gate} =$~15~V. The
marked regions (I)-(III) are investigated later on in greater
detail. The inset shows a micrograph of the device.} \label{BG and
SEM}
\end{figure}

Figure \ref{BG and SEM} shows the measured two-terminal conductance
$G$ through the device as a function of the back-gate voltage
$V_{gate}$, for two different temperatures, 230~mK and 1.5~K. At low
temperature, the Al contacts are in the super\-conducting state,
while for the higher temperature they are normal.
The charge neutrality point (CNP), indicated by the minimum in $G$,
lies at $V_{gate}=$~15~V which suggests p-doping. Charged impurities
introduce scattering and thus disorder \cite{Chen2008} which is
responsible for the broad shape of the dip in $G$.

At large doping, i.e. for small negative gate voltages, the
conductance $G$ is reduced in the normal state by 1~e$^{2}$/h, while
$G$ remains the same near the CNP. This can be explained by the
series resistance of the Al leads adding to the device in the normal
state. Figure \ref{BG and SEM} also shows that the CF are enhanced
in the super\-condcuting state. However, since the measurement in
this figure was performed with an applied source-drain bias voltage
$V_{sd}=$~140~$\mu$V which is similar to $\Delta$, the CF are not
fully enhanced by Andreev reflection.

We want to focus on three distinct regimes that can be identified
in figure \ref{BG and SEM}: (I) high doping with slow change in $G$,
(II) intermediate doping with steeply changing $G$ and (III) near
the CNP. We expect diffusive transport as in disordered metals in
regime (I), whereas in regime (III) near the CNP the graphene
properties should dominate.

\begin{figure}[htb]%.
\center
  \includegraphics{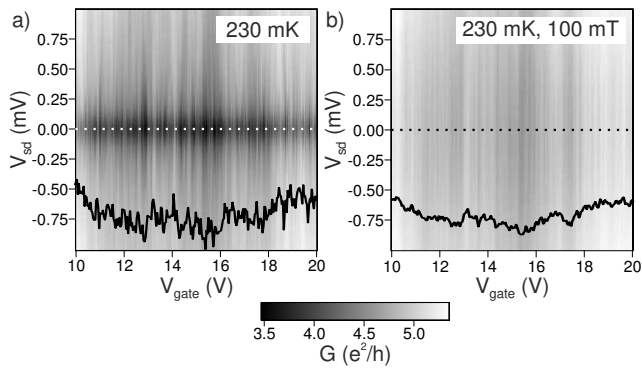}%
\caption{(a) Grey scale of the conductance $G$ as a function of the
gate voltage $V_{gate}$ and the applied source-drain bias $V_{sd}$
at 230~mK in region (III). The system enters the super\-conducting
gap for $|V_{sd}| <$~0.3~V. The conductance fluctuations in the gap
region increase, while outside the fluctuations are decreasing with
increased $|V_{sd}|$ . A cut in the super\-conducting gap (indicated
by the dotted line) is shown by the black curve. (b) Measurement of
$G$ with a 100~mT perpendicular magnetic field applied. The
conductance fluctuations are reduced, as can be seen from the cut
(black solid curve) at $V_{sd}=$~0~V.}
    \label{greyscales}
\end{figure}

Figure \ref{greyscales} displays grey scale plots of $G$ in region
(III)  near the CNP as a function of $V_{sd}$ and $V_{gate}$ at
230~mK. In figure \ref{greyscales}(a), the electrodes are in a
super\-conducting state. Due to the super\-conducting gap $\Delta$
transport is expected to be dominated by Andreev reflection at low
bias voltage $V_{sd} < 2 \Delta /e$, where the factor 2 accounts for
the two superconductors defining source and drain contacts. In this
bias window the device displays a suppression $G$, indicative for
weakly coupled contacts. The fluctuations in $G$ as a function of
$V_{gate}$ are largest at zero bias and are reduced by an increased
$V_{sd}$. When applying a small magnetic field, as shown in figure
\ref{greyscales}(b), the super\-conductivity is suppressed.
Consequently, the fluctuations of $G$ at zero bias are visibly
reduced.

In order to quantify the change in the conductance fluctuations
originating  from enhancement due to Andreev reflections, we
investigate the standard deviation of the conduction, $\delta G$ as
a function of the bias voltage $V_{sd}$, as shown in
\cite{Trbovic2010}.

\begin{figure}[htb]%.
\center
  \includegraphics{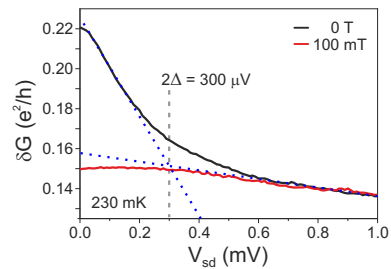}%
\caption{(online colour at: www.pss-b.com) Conductance fluctuations
$\delta G$ as a function of the applied bias voltage $V_{sd}$ at
230~mK in region (III). Two regimes can be observed in the
super\-conducting state (solid line): Outside of the
super\-conducting gap ($V_{sd}>$~0.3~mV), the fluctuations are
reduced in comparison to the values in the gap. In 100~mT
perpendicular field super\-conductivity is suppressed and $\delta G$
(dotted line). At $V_{sd} >$~0.6~mV the normal and super\-conducting
$\delta G$ are identical. Below $V_{sd}=$~0.3~mV, the
super\-conducting contacts enhance the fluctuations by a factor of
up to 1.5 when compared to $\delta G$ in the normal state at
100~mT.}
    \label{dG-junction}
\end{figure}

Figure \ref{dG-junction} shows a plot of $\delta G$ against $V_{sd}$
near the CNP in (III). When the electrodes are in a
super\-conducting state, the CF are largest at zero $V_{sd}$ with
$\delta G =$~0.22~e$^2$/h. With increasing $V_{sd}$ the CF decrease
and reach at $V_{sd}=$~1~mV a value of $\delta G =$~0.14~e$^2$/h. By
comparing the slopes of the steeply changing CF at small $V_{sd}$
and the more slowly varying CF at larger $V_{sd}$, we can extract a
crossover and thus a super\-conducting gap of $2 \Delta
\approx$~300~$\mu$V. In the normal conducting state at 100~mT, the
CF are about 0.14~e$^2$/h at $V_{sd}=$~1~mV, which coincides with
the value of the super\-conducting electrodes. However, for $V_{sd}
<$~0.3~mV the CF in the normal state saturate at $\delta
G=$~0.15~e$^2$/h. Thus the CF in the super\-conducting state are
enhanced by a factor of up to 1.5 compared to the normal state.

When comparing the normal state fluctuations $\delta G \approx$
0.15~e$^2$/h to the theoretical Altshuler-Lee-Stone (ALS) value of
$\delta G = 0.69 \sqrt{W/L}$~e$^2$/h $\approx$~1.2~e$^2$/h
\cite{Rycerz2007}, we find that the prediction is almost one order
of magnitude larger than the measured value. One way to resolve this
discrepancy is to assume that the phase coherence length $l_{\phi}$
is much shorter than the length $L$ of the device. Assuming that
each segment of length $l_{\phi}$ is fluctuating independently,
yields an overall reduced CF \cite{Trbovic2010}. Applying this
approach to the current values yields a phase coherence length of
300 nm. However, the fact that the measured CF is nearly doubled in
the super\-conducting states suggests that almost all electrons are
within the coherence length of one of the super\-conducting
interface, hence $l_{\phi} \gtrsim L/2$. This in strong
contradicting to the value of $l_{\phi}$ estimated from the normal
state fluctuations.

\begin{figure}[]%.
\center
  \includegraphics{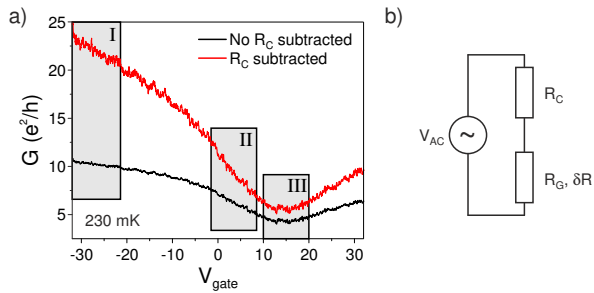}%
\caption{(online colour at: www.pss-b.com) (a) The conductance $G$
as a function  of the gate voltage $V_{gate}$ at 230~mK with and
without the subtraction of a constant series resistor $R_{C}$. When
$R_{C}$ is not subtracted (black curve) the CF are stronger in
region (III) than in region (I). With $R_{C}=$~1.35~k$\Omega$
subtracted the CF are stronger in region (I) and decrease when
approaching the CNP in region (III). (b) Circuit diagram of the
device. A constant resistor $R_{C}$ is in series with the graphene
resistor $R_{G}$. All fluctuations $\delta R$ originate in the
graphene part.}
    \label{minusrc}
\end{figure}

Another parameter which can reduce the apparent CF is a built-in
series resistance $R_{C}$. This is schematically shown in
\ref{minusrc}(b). The graphene part has the resistance $R_{G}$ which
fluctuates with $\delta R$, whereas $R_C$ is assumed to be constant.
In practice $R_{C}$ consists of the electrical lines in the
measurement set-up, the input impedance of the IV converter and the
resistance of the Al contact lines. In addition, it may also contain
to some extent the contact resistance of the device
\cite{Williams2009}. All this is intrinsic to two-terminal
measurements. We assume that $R_C$ is independent of the carrier
concentration in the graphene flake. The subtraction of $R_{C}$
corrects the measured fluctuations, $\delta G_{m}$, by $\delta G =
\left( 1+R_{C}/R_{G} \right)^{2} \cdot \delta G_{m}$. By adding the
known resistance of the measurement setup (measurement lines and IV
converter) we arrive at a lower boundary of $R_{C}
\approx$~1.35~k$\Omega$. An upper boundary of $R_{C} \approx
1.5$~k$\Omega$. can be extracted from the quantum Hall effect, which
due to the magnetic field measures the device in the normal state.
$R_C$ has been deduced from the shifts of the Hall plateaus.

Figure \ref{minusrc}(a) compares $G$ against $V_{gate}$ at 230~mK
with and  without $R_{C}=$~1.35~k$\Omega$ subtracted. As expected
the correction results in the largest change at large doping when
$G$ is large. The same is true for the CF, as we will show in the
following.

\begin{figure}[htb]%.
\center
  \includegraphics{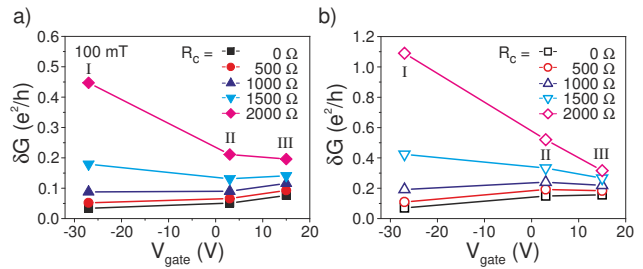}%
\caption{(online colour at: www.pss-b.com) Conductance fluctuations
$\delta G$ at 230~mK and $V_{sd}=$~0 in dependence on the gate
$V_{gate}$ and for different values of $R_{C}$. A B-spline was
subtracted from the original data to account for the changing
conductivity of the graphene. (a) Normal conducting electrodes at
100~mT. The effect of $R_{C}$ is strongest in region (I). Without
the subtraction of $R_{C}$, $\delta G$ increases from (I) to (III).
At the largest subtracted $R_{C}=$~2000~$\Omega$, $\delta G$
decreases from (I) to (III). (b) $\delta G$ in the super\-conducting
state. The CF are enhanced compared to the normal state. The dependence of the CF on $V_{gate}$ and $R_{C}$ follows the same tendency as in the normal state.}
    \label{dG}
\end{figure}

In order to assess the sensitivity of the CF on the value of $R_C$,
we have investigated the dependence of the CF measured at
$V_{sd}=$~0 and 230~mK for different $R_C$ values, ranging from 0 to
2~k$\Omega$ in regime (I) (high doping), (II) (intermediate doping),
and (III) (CNP). Figure \ref{dG} shows the result. In order to
calculate $\delta G$ a B-spline was subtracted to account for the
non-constant background of the graphene conductance $G(V_{gate})$.
The absolute values of the CF are lower than the values found in
figure \ref{dG-junction}, because there only a linear background was
removed.

In figure \ref{dG}(a) the electrical contacts are in a normal
conducting state due to the application of a 100~mT magnetic field.
The as measured data (no $R_{C}$ subtracted) shows the smallest CF
which increase from 0.03~e$^2$/h in (I) to 0.08~e$^2$/h in (III).
Around the series resistance that we estimated for our set-up
(1.35~k$\Omega$ - 1.5~k$\Omega$), the fluctuations are nearly
independent of $V_{gate}$, with values of $\sim$ 0.14~e$^2$/h in
regimes (I) and (III). If $R_{C}=$~2~k$\Omega$ is subtracted the CF
in (I) reach 0.45~e$^2$/h and are reduced to 0.2~e$^2$/h in (III).

Without the subtraction of $R_{C}$, the CF are minimal at large
doping and increase near the CNP. When $R_{C}$ is subtracted the CF
increase. However, they increase the most in region (I), less strong
in region (II) and the weakest increase is in region (III). This can
lead to a reversal of the dependence of the CFs on the $V_{gate}$.

The conductance fluctuations $\delta G$ with super\-conducting
contacts  are plotted in figure \ref{dG}(b). The measured data
without any $R_{C}$ subtracted shows an increase from 0.07~e$^2$/h
in (I) to 0.16~e$^2$/h near the CNP in (III). Again, for values of
$R_{C}$ that lie within the estimates for our set-up, the CF vary
only weakly with $V_{gate}$ from 0.32~e$^2$/h in (I) to 0.25~e$^2$/h
in (III). At higher values of $R_{C}$ the CF decrease significantly
from (I) to (III).

Evidently, the super\-conductivity enhances the CF in all three
regimes.  In (I) and (III) the enhancement lies close to a factor of
2. Yet in regime (II) it appears to be consistently higher and
reaches a factor of 3. In part this large enhancement factor
originates from the resistance of the aluminium electrodes, which is
approximately 100~$\Omega$ in the normal state in our device. If we
consider this difference, we find an enhancement factor of around
1.9 for (I) and (III), but still above 2 for regime (II).

In recent experiments contradicting observations were made.  In
\cite{Ojeda-Aristizabal2010} an increase of the CF near the CNP was
seen, where the increase of the CF from high charge carrier
concentration to near the CNP was attributed to the formation of
electron-hole puddles. In contrast, a decrease of CF near the CNP
was found in \cite{Staley2008,Chen2010}. The changes in the CF were
reported to be of a factor of 3 or larger. Gate independent CF were
observed in \cite{Horsell2009}. As in our device,
\cite{Ojeda-Aristizabal2010} rely on two-terminal measurements,
which include $R_{C}$, whereas \cite{Staley2008,Chen2010} use a
four-terminal set-up, where $R_{C}$ is irrelevant. As shown above,
we can see all three gate dependences if we subtract $R_{C}$.
Nevertheless, for the range of $R_{C}$ that we estimate for our
set-up, we find nearly no change in CF at high charge carrier
density and near the CNP in our graphene device.

%-------------------------------------

\section{Conclusions}

Using graphene devices with super\-conducting contacts, we have
shown that CFs are enhanced in the super\-conducting relative to the
normal state of the contacts. This enhancement amounts to a factor
between 1.4 and 2 and thus suggests that all electrons in the device
of length $L=$~1.3~$\mu$m are within the coherence length of one of
the two contacts where Andreev reflection occurs which is the origin
of the doubling \cite{Beenakker1995}. The phase coherence length at
230~mK is therefore large and amounts to $l_{\phi} \gtrsim L/2$. We
established a constant series resistor $R_{C}$ to account for our
two-terminal measurement set-up and device and investigated its
impact on the CF. For values of $R_{C}$ that we estimated for our
set-up and device the CF remain smaller than e$^2$/h. Furthermore,
we investigated the dependence of the CF on the back-gate voltage
$V_{gate}$. A strong dependence on $R_{C}$ is found. Without
subtraction of $R_{C}$ the CF increase around the CNP compared to
the CF at high doping. In the range of appropriate $R_{C}$ the CF
are nearly independent of $V_{gate}$.

\begin{acknowledgements}
This work is financially supported by the NCCR on Nanoscale Science,
the  NCCR on Quantum Science, the Swiss NSF and EU-FP7-IST project
SE2ND. We are grateful to Markus Weiss and Romain Maurand for
discussions.
\end{acknowledgements}

% Use the following code if you wish to generate your bibliography with BibTeX;
% replace the string "pss-demo" below with the name(s) of
% the BibTeX data base(s) you want to use.
% The resulting bibliography-output (the content of the .bbl file)
% must be pasted back into this file before submission.
% Please also include your BibTeX data base file(s) in your submission
% so that we can re-run BibTeX if necessary.
%
%\bibliographystyle{pss}
%\bibliography{pss-demo}

\begin{thebibliography}{19}%
\makeatletter
\providecommand \@ifxundefined [1]{%
 \@ifx{#1\undefined}
}%
\providecommand \@ifnum [1]{%
 \ifnum #1\expandafter \@firstoftwo
 \else \expandafter \@secondoftwo
 \fi
}%
\providecommand \@ifx [1]{%
 \ifx #1\expandafter \@firstoftwo
 \else \expandafter \@secondoftwo
 \fi
}%
\providecommand \natexlab [1]{#1}%
\providecommand \enquote  [1]{``#1''}%
\providecommand \bibnamefont  [1]{#1}%
\providecommand \bibfnamefont [1]{#1}%
\providecommand \citenamefont [1]{#1}%
\providecommand \href@noop [0]{\@secondoftwo}%
\providecommand \href [0]{\begingroup \@sanitize@url \@href}%
\providecommand \@href[1]{\@@startlink{#1}\@@href}%
\providecommand \@@href[1]{\endgroup#1\@@endlink}%
\providecommand \@sanitize@url [0]{\catcode `\\12\catcode `\$12\catcode
  `\&12\catcode `\#12\catcode `\^12\catcode `\_12\catcode `\%12\relax}%
\providecommand \@@startlink[1]{}%
\providecommand \@@endlink[0]{}%
\providecommand \url  [0]{\begingroup\@sanitize@url \@url }%
\providecommand \@url [1]{\endgroup\@href {#1}{\urlprefix }}%
\providecommand \urlprefix  [0]{URL }%
\providecommand \Eprint [0]{\href }%
\providecommand \doibase [0]{http://dx.doi.org/}%
\providecommand \selectlanguage [0]{\@gobble}%
\providecommand \bibinfo  [0]{\@secondoftwo}%
\providecommand \bibfield  [0]{\@secondoftwo}%
\providecommand \translation [1]{[#1]}%
\providecommand \BibitemOpen [0]{}%
\providecommand \bibitemStop [0]{}%
\providecommand \bibitemNoStop [0]{.\EOS\space}%
\providecommand \EOS [0]{\spacefactor3000\relax}%
\providecommand \BibitemShut  [1]{\csname bibitem#1\endcsname}%
\let\auto@bib@innerbib\@empty
%</preamble>
\bibitem [{\citenamefont {Morozov}\ \emph {et~al.}(2006)\citenamefont
  {Morozov}, \citenamefont {Novoselov}, \citenamefont {Katsnelson},
  \citenamefont {Schedin}, \citenamefont {Ponomarenko}, \citenamefont {Jiang},\
  and\ \citenamefont {Geim}}]{Morozov2006}%
  \BibitemOpen
  \bibfield  {author} {\bibinfo {author} {\bibfnamefont {S.}~\bibnamefont
  {Morozov}}, \bibinfo {author} {\bibfnamefont {K.}~\bibnamefont {Novoselov}},
  \bibinfo {author} {\bibfnamefont {M.}~\bibnamefont {Katsnelson}}, \bibinfo
  {author} {\bibfnamefont {F.}~\bibnamefont {Schedin}}, \bibinfo {author}
  {\bibfnamefont {L.}~\bibnamefont {Ponomarenko}}, \bibinfo {author}
  {\bibfnamefont {D.}~\bibnamefont {Jiang}}, \ and\ \bibinfo {author}
  {\bibfnamefont {A.}~\bibnamefont {Geim}},\ }\href@noop {} {\bibfield
  {journal} {\bibinfo  {journal} {Physical Review Letters}\ }\textbf {\bibinfo
  {volume} {97}},\ \bibinfo {pages} {16801} (\bibinfo {year}
  {2006})}\BibitemShut {NoStop}%
\bibitem [{\citenamefont {Heersche}\ \emph {et~al.}(2007)\citenamefont
  {Heersche}, \citenamefont {Jarillo-Herrero}, \citenamefont {Oostinga},
  \citenamefont {Vandersypen},\ and\ \citenamefont {Morpurgo}}]{Heersche2007}%
  \BibitemOpen
  \bibfield  {author} {\bibinfo {author} {\bibfnamefont {H.~B.}\ \bibnamefont
  {Heersche}}, \bibinfo {author} {\bibfnamefont {P.}~\bibnamefont
  {Jarillo-Herrero}}, \bibinfo {author} {\bibfnamefont {J.~B.}\ \bibnamefont
  {Oostinga}}, \bibinfo {author} {\bibfnamefont {L.~M.}\ \bibnamefont
  {Vandersypen}}, \ and\ \bibinfo {author} {\bibfnamefont {A.~F.}\ \bibnamefont
  {Morpurgo}},\ }\href {http://dx.doi.org/10.1140/epjst/e2007-00223-7}
  {\bibfield  {journal} {\bibinfo  {journal} {The European Physical Journal -
  Special Topics}\ }\textbf {\bibinfo {volume} {148}},\ \bibinfo {pages} {27}
  (\bibinfo {year} {2007})},\ \bibinfo {note}
  {10.1140/epjst/e2007-00223-7}\BibitemShut {NoStop}%
\bibitem [{\citenamefont {Ojeda-Aristizabal}\ \emph {et~al.}(2009)\citenamefont
  {Ojeda-Aristizabal}, \citenamefont {Ferrier}, \citenamefont {Gueron},\ and\
  \citenamefont {Bouchiat}}]{Ojeda-Aristizabal2009}%
  \BibitemOpen
  \bibfield  {author} {\bibinfo {author} {\bibfnamefont {C.}~\bibnamefont
  {Ojeda-Aristizabal}}, \bibinfo {author} {\bibfnamefont {M.}~\bibnamefont
  {Ferrier}}, \bibinfo {author} {\bibfnamefont {S.}~\bibnamefont {Gueron}}, \
  and\ \bibinfo {author} {\bibfnamefont {H.}~\bibnamefont {Bouchiat}},\ }\href
  {\doibase 10.1103/PhysRevB.79.165436} {\bibfield  {journal} {\bibinfo
  {journal} {Phys. Rev. B}\ }\textbf {\bibinfo {volume} {79}},\ \bibinfo
  {pages} {165436} (\bibinfo {year} {2009})}\BibitemShut {NoStop}%
\bibitem [{\citenamefont {Sk\'akalov\'a}\ \emph {et~al.}(2009)\citenamefont
  {Sk\'akalov\'a}, \citenamefont {Kaiser}, \citenamefont {Yoo}, \citenamefont
  {Obergfell},\ and\ \citenamefont {Roth}}]{Skakalova2009}%
  \BibitemOpen
  \bibfield  {author} {\bibinfo {author} {\bibfnamefont {V.}~\bibnamefont
  {Sk\'akalov\'a}}, \bibinfo {author} {\bibfnamefont {A.~B.}\ \bibnamefont
  {Kaiser}}, \bibinfo {author} {\bibfnamefont {J.~S.}\ \bibnamefont {Yoo}},
  \bibinfo {author} {\bibfnamefont {D.}~\bibnamefont {Obergfell}}, \ and\
  \bibinfo {author} {\bibfnamefont {S.}~\bibnamefont {Roth}},\ }\href {\doibase
  10.1103/PhysRevB.80.153404} {\bibfield  {journal} {\bibinfo  {journal} {Phys.
  Rev. B}\ }\textbf {\bibinfo {volume} {80}},\ \bibinfo {pages} {153404}
  (\bibinfo {year} {2009})}\BibitemShut {NoStop}%
\bibitem [{\citenamefont {Trbovic}\ \emph {et~al.}(2010)\citenamefont
  {Trbovic}, \citenamefont {Minder}, \citenamefont {Freitag},\ and\
  \citenamefont {Sch\"onenberger}}]{Trbovic2010}%
  \BibitemOpen
  \bibfield  {author} {\bibinfo {author} {\bibfnamefont {J.}~\bibnamefont
  {Trbovic}}, \bibinfo {author} {\bibfnamefont {N.}~\bibnamefont {Minder}},
  \bibinfo {author} {\bibfnamefont {F.}~\bibnamefont {Freitag}}, \ and\
  \bibinfo {author} {\bibfnamefont {C.}~\bibnamefont {Sch\"onenberger}},\
  }\href@noop {} {\bibfield  {journal} {\bibinfo  {journal} {Nanotechnology}\
  }\textbf {\bibinfo {volume} {21}},\ \bibinfo {pages} {274005 (8 pp.)}
  (\bibinfo {year} {2010})}\BibitemShut {NoStop}%
\bibitem [{\citenamefont {Geim}\ and\ \citenamefont
  {Novoselov}(2007)}]{Geim2007}%
  \BibitemOpen
  \bibfield  {author} {\bibinfo {author} {\bibfnamefont {A.~K.}\ \bibnamefont
  {Geim}}\ and\ \bibinfo {author} {\bibfnamefont {K.~S.}\ \bibnamefont
  {Novoselov}},\ }\href {http://dx.doi.org/10.1038/nmat1849} {\bibfield
  {journal} {\bibinfo  {journal} {Nat Mater}\ }\textbf {\bibinfo {volume}
  {6}},\ \bibinfo {pages} {183} (\bibinfo {year} {2007})}\BibitemShut {NoStop}%
\bibitem [{\citenamefont {McCann}\ \emph {et~al.}(2006)\citenamefont {McCann},
  \citenamefont {Kechedzhi}, \citenamefont {Fal'ko}, \citenamefont {Suzuura},
  \citenamefont {Ando},\ and\ \citenamefont {Altshuler}}]{McCann2006}%
  \BibitemOpen
  \bibfield  {author} {\bibinfo {author} {\bibfnamefont {E.}~\bibnamefont
  {McCann}}, \bibinfo {author} {\bibfnamefont {K.}~\bibnamefont {Kechedzhi}},
  \bibinfo {author} {\bibfnamefont {V.~I.}\ \bibnamefont {Fal'ko}}, \bibinfo
  {author} {\bibfnamefont {H.}~\bibnamefont {Suzuura}}, \bibinfo {author}
  {\bibfnamefont {T.}~\bibnamefont {Ando}}, \ and\ \bibinfo {author}
  {\bibfnamefont {B.~L.}\ \bibnamefont {Altshuler}},\ }\href {\doibase
  10.1103/PhysRevLett.97.146805} {\bibfield  {journal} {\bibinfo  {journal}
  {Phys. Rev. Lett.}\ }\textbf {\bibinfo {volume} {97}},\ \bibinfo {pages}
  {146805} (\bibinfo {year} {2006})}\BibitemShut {NoStop}%
\bibitem [{\citenamefont {Rycerz}\ \emph {et~al.}(2007)\citenamefont {Rycerz},
  \citenamefont {Tworzydlo},\ and\ \citenamefont {Beenakker}}]{Rycerz2007}%
  \BibitemOpen
  \bibfield  {author} {\bibinfo {author} {\bibfnamefont {A.}~\bibnamefont
  {Rycerz}}, \bibinfo {author} {\bibfnamefont {J.}~\bibnamefont {Tworzydlo}}, \
  and\ \bibinfo {author} {\bibfnamefont {C.~W.~J.}\ \bibnamefont {Beenakker}},\
  }\href {\doibase 10.1209/0295-5075/79/57003} {\bibfield  {journal} {\bibinfo
  {journal} {EPL}\ }\textbf {\bibinfo {volume} {79}},\ \bibinfo {pages} {57003}
  (\bibinfo {year} {2007})}\BibitemShut {NoStop}%
\bibitem [{\citenamefont {Borunda}\ \emph {et~al.}(2011)\citenamefont
  {Borunda}, \citenamefont {Berezovsky}, \citenamefont {Westervelt},\ and\
  \citenamefont {Heller}}]{Borunda2011}%
  \BibitemOpen
  \bibfield  {author} {\bibinfo {author} {\bibfnamefont {M.~F.}\ \bibnamefont
  {Borunda}}, \bibinfo {author} {\bibfnamefont {J.}~\bibnamefont {Berezovsky}},
  \bibinfo {author} {\bibfnamefont {R.~M.}\ \bibnamefont {Westervelt}}, \ and\
  \bibinfo {author} {\bibfnamefont {E.~J.}\ \bibnamefont {Heller}},\ }\href
  {\doibase 10.1021/nn103450d} {\bibfield  {journal} {\bibinfo  {journal} {ACS
  Nano}\ }\textbf {\bibinfo {volume} {5}},\ \bibinfo {pages} {3622} (\bibinfo
  {year} {2011})},\ \Eprint
  {http://arxiv.org/abs/http://pubs.acs.org/doi/pdf/10.1021/nn103450d}
  {http://pubs.acs.org/doi/pdf/10.1021/nn103450d} \BibitemShut {NoStop}%
\bibitem [{\citenamefont {Kharitonov}\ and\ \citenamefont
  {Efetov}(2008)}]{Kharitonov2008}%
  \BibitemOpen
  \bibfield  {author} {\bibinfo {author} {\bibfnamefont {M.~Y.}\ \bibnamefont
  {Kharitonov}}\ and\ \bibinfo {author} {\bibfnamefont {K.~B.}\ \bibnamefont
  {Efetov}},\ }\href {\doibase 10.1103/PhysRevB.78.033404} {\bibfield
  {journal} {\bibinfo  {journal} {Phys. Rev. B}\ }\textbf {\bibinfo {volume}
  {78}},\ \bibinfo {pages} {033404} (\bibinfo {year} {2008})}\BibitemShut
  {NoStop}%
\bibitem [{\citenamefont {Beenakker}(1995)}]{Beenakker1995}%
  \BibitemOpen
  \bibfield  {author} {\bibinfo {author} {\bibfnamefont {C.~W.~J.}\
  \bibnamefont {Beenakker}},\ }\href@noop {} {\emph {\bibinfo {title}
  {Mesoscopic Quantum Physics}}},\ edited by\ \bibinfo {editor} {\bibfnamefont
  {E.}~\bibnamefont {Akkermans}}, \bibinfo {editor} {\bibfnamefont
  {G.}~\bibnamefont {Montambaux}}, \bibinfo {editor} {\bibfnamefont {J.-L.}\
  \bibnamefont {Pichard}}, \ and\ \bibinfo {editor} {\bibfnamefont
  {J.}~\bibnamefont {Zin-Justin}}\ (\bibinfo {year} {1995})\BibitemShut
  {NoStop}%
\bibitem [{\citenamefont {Novoselov}\ \emph {et~al.}(2005)\citenamefont
  {Novoselov}, \citenamefont {Jiang}, \citenamefont {Schedin}, \citenamefont
  {Booth}, \citenamefont {Khotkevich}, \citenamefont {Morozov},\ and\
  \citenamefont {Geim}}]{Novoselov2005}%
  \BibitemOpen
  \bibfield  {author} {\bibinfo {author} {\bibfnamefont {K.}~\bibnamefont
  {Novoselov}}, \bibinfo {author} {\bibfnamefont {D.}~\bibnamefont {Jiang}},
  \bibinfo {author} {\bibfnamefont {F.}~\bibnamefont {Schedin}}, \bibinfo
  {author} {\bibfnamefont {T.}~\bibnamefont {Booth}}, \bibinfo {author}
  {\bibfnamefont {V.}~\bibnamefont {Khotkevich}}, \bibinfo {author}
  {\bibfnamefont {S.}~\bibnamefont {Morozov}}, \ and\ \bibinfo {author}
  {\bibfnamefont {A.}~\bibnamefont {Geim}},\ }\href@noop {} {\bibfield
  {journal} {\bibinfo  {journal} {Proc. Natl. Acad. Sci. U.S.A.}\ }\textbf
  {\bibinfo {volume} {102}},\ \bibinfo {pages} {10451} (\bibinfo {year}
  {2005})}\BibitemShut {NoStop}%
\bibitem [{\citenamefont {Ferrari}\ \emph {et~al.}(2006)\citenamefont
  {Ferrari}, \citenamefont {Meyer}, \citenamefont {Scardaci}, \citenamefont
  {Casiraghi}, \citenamefont {Lazzeri}, \citenamefont {Mauri}, \citenamefont
  {Piscanec}, \citenamefont {Jiang}, \citenamefont {Novoselov}, \citenamefont
  {Roth},\ and\ \citenamefont {Geim}}]{Ferrari2006}%
  \BibitemOpen
  \bibfield  {author} {\bibinfo {author} {\bibfnamefont {A.~C.}\ \bibnamefont
  {Ferrari}}, \bibinfo {author} {\bibfnamefont {J.~C.}\ \bibnamefont {Meyer}},
  \bibinfo {author} {\bibfnamefont {V.}~\bibnamefont {Scardaci}}, \bibinfo
  {author} {\bibfnamefont {C.}~\bibnamefont {Casiraghi}}, \bibinfo {author}
  {\bibfnamefont {M.}~\bibnamefont {Lazzeri}}, \bibinfo {author} {\bibfnamefont
  {F.}~\bibnamefont {Mauri}}, \bibinfo {author} {\bibfnamefont
  {S.}~\bibnamefont {Piscanec}}, \bibinfo {author} {\bibfnamefont
  {D.}~\bibnamefont {Jiang}}, \bibinfo {author} {\bibfnamefont {K.~S.}\
  \bibnamefont {Novoselov}}, \bibinfo {author} {\bibfnamefont {S.}~\bibnamefont
  {Roth}}, \ and\ \bibinfo {author} {\bibfnamefont {A.~K.}\ \bibnamefont
  {Geim}},\ }\href {\doibase 10.1103/PhysRevLett.97.187401} {\bibfield
  {journal} {\bibinfo  {journal} {Phys. Rev. Lett.}\ }\textbf {\bibinfo
  {volume} {97}},\ \bibinfo {pages} {187401} (\bibinfo {year}
  {2006})}\BibitemShut {NoStop}%
\bibitem [{\citenamefont {Chen}\ \emph {et~al.}(2008)\citenamefont {Chen},
  \citenamefont {Jang}, \citenamefont {Adam}, \citenamefont {Fuhrer},
  \citenamefont {Williams},\ and\ \citenamefont {Ishigami}}]{Chen2008}%
  \BibitemOpen
  \bibfield  {author} {\bibinfo {author} {\bibfnamefont {J.-H.}\ \bibnamefont
  {Chen}}, \bibinfo {author} {\bibfnamefont {C.}~\bibnamefont {Jang}}, \bibinfo
  {author} {\bibfnamefont {S.}~\bibnamefont {Adam}}, \bibinfo {author}
  {\bibfnamefont {M.~S.}\ \bibnamefont {Fuhrer}}, \bibinfo {author}
  {\bibfnamefont {E.~D.}\ \bibnamefont {Williams}}, \ and\ \bibinfo {author}
  {\bibfnamefont {M.}~\bibnamefont {Ishigami}},\ }\href
  {http://dx.doi.org/10.1038/nphys935} {\bibfield  {journal} {\bibinfo
  {journal} {Nat Phys}\ }\textbf {\bibinfo {volume} {4}},\ \bibinfo {pages}
  {377} (\bibinfo {year} {2008})}\BibitemShut {NoStop}%
\bibitem [{\citenamefont {Williams}\ \emph {et~al.}(2009)\citenamefont
  {Williams}, \citenamefont {Abanin}, \citenamefont {DiCarlo}, \citenamefont
  {Levitov},\ and\ \citenamefont {Marcus}}]{Williams2009}%
  \BibitemOpen
  \bibfield  {author} {\bibinfo {author} {\bibfnamefont {J.~R.}\ \bibnamefont
  {Williams}}, \bibinfo {author} {\bibfnamefont {D.~A.}\ \bibnamefont
  {Abanin}}, \bibinfo {author} {\bibfnamefont {L.}~\bibnamefont {DiCarlo}},
  \bibinfo {author} {\bibfnamefont {L.~S.}\ \bibnamefont {Levitov}}, \ and\
  \bibinfo {author} {\bibfnamefont {C.~M.}\ \bibnamefont {Marcus}},\ }\href
  {\doibase 10.1103/PhysRevB.80.045408} {\bibfield  {journal} {\bibinfo
  {journal} {Phys. Rev. B}\ }\textbf {\bibinfo {volume} {80}},\ \bibinfo
  {pages} {045408} (\bibinfo {year} {2009})}\BibitemShut {NoStop}%
\bibitem [{\citenamefont {Ojeda-Aristizabal}\ \emph {et~al.}(2010)\citenamefont
  {Ojeda-Aristizabal}, \citenamefont {Monteverde}, \citenamefont {Weil},
  \citenamefont {Ferrier}, \citenamefont {Gueron},\ and\ \citenamefont
  {Bouchiat}}]{Ojeda-Aristizabal2010}%
  \BibitemOpen
  \bibfield  {author} {\bibinfo {author} {\bibfnamefont {C.}~\bibnamefont
  {Ojeda-Aristizabal}}, \bibinfo {author} {\bibfnamefont {M.}~\bibnamefont
  {Monteverde}}, \bibinfo {author} {\bibfnamefont {R.}~\bibnamefont {Weil}},
  \bibinfo {author} {\bibfnamefont {M.}~\bibnamefont {Ferrier}}, \bibinfo
  {author} {\bibfnamefont {S.}~\bibnamefont {Gueron}}, \ and\ \bibinfo {author}
  {\bibfnamefont {H.}~\bibnamefont {Bouchiat}},\ }\href {\doibase
  10.1103/PhysRevLett.104.186802} {\bibfield  {journal} {\bibinfo  {journal}
  {Phys. Rev. Lett.}\ }\textbf {\bibinfo {volume} {104}},\ \bibinfo {pages}
  {186802} (\bibinfo {year} {2010})}\BibitemShut {NoStop}%
\bibitem [{\citenamefont {Staley}\ \emph {et~al.}(2008)\citenamefont {Staley},
  \citenamefont {Puls},\ and\ \citenamefont {Liu}}]{Staley2008}%
  \BibitemOpen
  \bibfield  {author} {\bibinfo {author} {\bibfnamefont {N.}~\bibnamefont
  {Staley}}, \bibinfo {author} {\bibfnamefont {C.}~\bibnamefont {Puls}}, \ and\
  \bibinfo {author} {\bibfnamefont {Y.}~\bibnamefont {Liu}},\ }\href@noop {}
  {\bibfield  {journal} {\bibinfo  {journal} {Phys. Rev. B}\ }\textbf {\bibinfo
  {volume} {77}},\ \bibinfo {pages} {155429} (\bibinfo {year}
  {2008})}\BibitemShut {NoStop}%
\bibitem [{\citenamefont {Chen}\ \emph {et~al.}(2010)\citenamefont {Chen},
  \citenamefont {Bae}, \citenamefont {Chialvo}, \citenamefont {Dirks},
  \citenamefont {Bezryadin},\ and\ \citenamefont {Mason}}]{Chen2010}%
  \BibitemOpen
  \bibfield  {author} {\bibinfo {author} {\bibfnamefont {Y.-F.}\ \bibnamefont
  {Chen}}, \bibinfo {author} {\bibfnamefont {M.-H.}\ \bibnamefont {Bae}},
  \bibinfo {author} {\bibfnamefont {C.}~\bibnamefont {Chialvo}}, \bibinfo
  {author} {\bibfnamefont {T.}~\bibnamefont {Dirks}}, \bibinfo {author}
  {\bibfnamefont {A.}~\bibnamefont {Bezryadin}}, \ and\ \bibinfo {author}
  {\bibfnamefont {N.}~\bibnamefont {Mason}},\ }\href
  {http://stacks.iop.org/0953-8984/22/i=20/a=205301} {\bibfield  {journal}
  {\bibinfo  {journal} {Journal of Physics: Condensed Matter}\ }\textbf
  {\bibinfo {volume} {22}},\ \bibinfo {pages} {205301} (\bibinfo {year}
  {2010})}\BibitemShut {NoStop}%
\bibitem [{\citenamefont {Horsell}\ \emph {et~al.}(2009)\citenamefont
  {Horsell}, \citenamefont {Savchenko}, \citenamefont {Tikhonenko},
  \citenamefont {Kechedzhi}, \citenamefont {Lerner},\ and\ \citenamefont
  {Fal'ko}}]{Horsell2009}%
  \BibitemOpen
  \bibfield  {author} {\bibinfo {author} {\bibfnamefont {D.}~\bibnamefont
  {Horsell}}, \bibinfo {author} {\bibfnamefont {A.}~\bibnamefont {Savchenko}},
  \bibinfo {author} {\bibfnamefont {F.}~\bibnamefont {Tikhonenko}}, \bibinfo
  {author} {\bibfnamefont {K.}~\bibnamefont {Kechedzhi}}, \bibinfo {author}
  {\bibfnamefont {I.}~\bibnamefont {Lerner}}, \ and\ \bibinfo {author}
  {\bibfnamefont {V.}~\bibnamefont {Fal'ko}},\ }\href {\doibase DOI:
  10.1016/j.ssc.2009.02.058} {\bibfield  {journal} {\bibinfo  {journal} {Solid
  State Communications}\ }\textbf {\bibinfo {volume} {149}},\ \bibinfo {pages}
  {1041 } (\bibinfo {year} {2009})},\ \bibinfo {note} {recent Progress in
  Graphene Studies}\BibitemShut {NoStop}%
\end{thebibliography}
%
% Replace the following example bibliography with your references
% before submission:


\end{document}